\documentclass[journal]{IEEEtran}
\usepackage{graphicx}
\usepackage{xcolor}
\usepackage{url}

\usepackage[compress]{cite}
\usepackage{psfrag}
\usepackage[font=footnotesize]{subfig}
\usepackage{multirow,multicol}
\usepackage[cmex10]{amsmath}
\usepackage{amsfonts}
\usepackage{amssymb}
\usepackage{verbatim}
\usepackage{pifont}

\usepackage{tikz-timing}
\usetikztiminglibrary[rising arrows]{clockarrows}

\newcommand{\cA}{\mathcal{A}} 
\newcommand{\cB}{\mathcal{B}} 
\newcommand{\cC}{\mathcal{C}} 
\newcommand{\cX}{\mathcal{X}} 

\newcommand{\RE}{\texttiming[timing/c/rising arrows, timing/c/arrow pos=.7]{2{C}}}
\newcommand{\FE}{\texttiming[timing/c/falling arrows, timing/c/arrow pos=.7]{1{HC}}}
\newcommand{\RBT}[1]{\bf{\textcolor{red}{#1}}}
\newcommand{\OD}{$\textcolor{red}{\boldsymbol{V_{th}\downarrow}}$}
\newcommand{\OU}{$\textcolor{red}{\boldsymbol{V_{th}\uparrow}}$}

\begin{document}

\title{
Impact of Sampler Offset on Jitter Transfer in Clock and Data Recovery Circuits}
\author{
\IEEEauthorblockN{Naveen Kadayinti, Maryam Shojaei Baghini and Dinesh K. Sharma}
\thanks{
Naveen Kadayinti is with the Dept. of Electrical Engineering at Indian Institute of Technology 
Dharwad, Karnataka, India. Maryam Shojaei Baghini and Dinesh K. Sharma are 
with the Dept. of Electrical Engineering at Indian Institute of Technology Bombay, Mumbai, India. }
\thanks{(emails : naveen@iitdh.ac.in, mshojaei@ee.iitb.ac.in, dinesh@ee.iitb.ac.in)}
}

\maketitle
\begin{abstract}
This paper shows how the input offset of sampling flip-flops in the 
Alexander phase detector affects the jitter transfer from data to the recovered clock 
in a clock data recovery circuit. 
The Alexander phase detector samples the data at both the edges of the clock in order 
to recover the data, as well as the clock timing information. 
The timing information is used in a clock recovery circuit, which is basically a 
PLL or a DLL. Once the PLL (or DLL) is locked, the phase detector samples the data 
at the center of the eye as well as at the data transitions. It is shown how the 
offset of the sampling flip-flop that samples the data at its transitions influences the 
jitter transfer from data to 
the recovered clock. Importantly, it is shown that zero offset is not always the best case.
The effect is studied for different levels of data dependent 
jitter. The mechanism of this phenomenon is explained and the predictions are supported 
with simulations. The paper also discusses a tracking circuit that keeps the offset at 
the minimum jitter point. 
\end{abstract}


\section{Introduction}
\label{sec:intro}
Clock data recovery circuits, which are essential in receivers of 
high speed serial links, have stringent jitter performance 
requirements. This is due to the fact that the jitter
performance of these circuits decides the overall timing margines 
and Bit Error Rates (BER) of the systems. With increasing data rates 
the available absolute timing margines are reducing. Designers spend 
a lot of effort in analyzing serial links to ensure their jitter 
and BER performance is satisfactory~\cite{kundert_CDR}.
A clock data recovery circuit comprises of a phase detector, which 
senses the phase difference between the clock and the data. The error 
output from the phase detector is used in a negative feedback loop through a 
voltage controlled oscillator (or a phase rotator) 
to generate the correct sampling clock. 
In such a phase locked system, there are two main sources of jitter, one being random 
jitter due to thermal noise and the second being deterministic jitter due to 
inter-symbol-interference (ISI)~\cite{ddj_prediction}. 
The authors of~\cite{jitter-ber} have analyzed the jitter generation 
mechanism and proposed techniques of compensating it via equalization. 
In this paper, we show how the offsets in the sampling flip-flops of the phase detectors 
determine the jitter transfer from data to recovered clock. We show that 
in the presence of ISI, the minimum jitter in the recovered clock is \emph{not} when the 
sampling flip-flops have zero offset. We analyse the system to explain why this is 
the case and design a circuit that recovers the optimum sampling threshold along with 
the clock phase.

The paper is organized as follows. Section~\ref{sec:alex} discusses the working of the bang-bang 
phase detector in the presence of jitter in the data. Section~\ref{sec:jitteroffset} discusses the 
working of clock and data recovery circuits for channels with different bandwidths. 
A circuit for recovering optimum sampling threshold is shown in Section~\ref{sec:auto} and 
the conclusions are presented in Section~\ref{sec:conclusion}.

\section{Working of the Alexander phase detector in the presence of jitter in the
data} 
\label{sec:alex}
\begin{figure}[h!] 
\centering 
\psfrag{Data}{\small{Data}}
\psfrag{Clock}{\small{Clock}} 
\psfrag{DN}{\small{DN}} 
\psfrag{UP}{\small{UP}}
\psfrag{D}{\small{D}} 
\psfrag{Q}{\small{Q}} 
\psfrag{D1}{\small{$F_1$}}
\psfrag{D2}{\small{$F_2$}} 
\psfrag{D3}{\small{$F_3$}} 
\psfrag{D4}{\small{$F_4$}} 
\includegraphics[width=7cm]{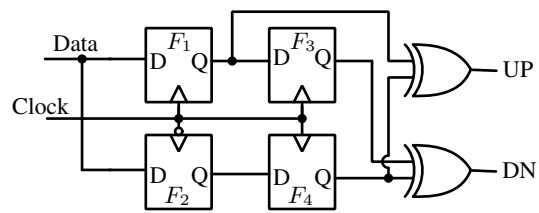}
\caption{Circuit diagram of the Alexander phase detector.} 
\label{fig:alex_pd}
\end{figure}

\begin{figure}[h!] 
\centering 
\psfrag{Clock Early}{\small{Clock Early}}
\psfrag{Clock Late}{\small{Clock Late}} 
\psfrag{Clock Locked}{\small{Clock Locked}}  
\psfrag{Data}{\small{Data}} 
\psfrag{Clock}{\small{Clock}}
\psfrag{A}{\small{A}} 
\psfrag{B}{\small{B}} 
\psfrag{C}{\small{C}}
\includegraphics[width=8cm]{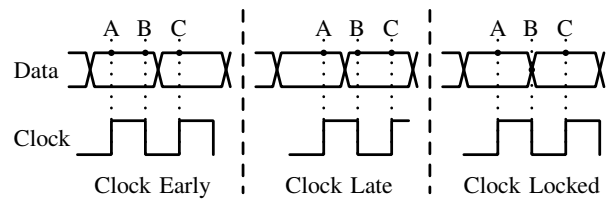} 
\caption{Timing diagram of sampling instants of an Alexander phase detector
under different possible phase differences between the clock and the data.} 
\label{fig:alex_pd_timing} 
\end{figure}

One of the most commonly used phase detector for sensing the phase
difference between clock and data is the Alexander bang-bang phase
detector~\cite{alexander}, which is shown in Fig.~\ref{fig:alex_pd}. 
The bang-bang phase detector samples the data at both the transitions of the clock to
determine the relative timing between the clock and data. The output of the phase detector 
is then used in a phase locked loop to generate a clock which is synchronous to the data.
Fig.~\ref{fig:alex_pd_timing} illustrates the sampling instants of the phase
detector for different possible phase differences between the clock and the data when the data signal 
has no jitter. 
%

Around every data transition, if the clock arrives early, sample `A' and `B' 
resolve to the same value and `C' resolves to a different value, and the clock is 
delayed for correction. Similarly, if the 
clock arrives late, sample `B' and `C' resolve to the same value and `A' resolves to 
a different value and the clock is advanced for correction. %
When the clock recovery loop settles, data is sampled at its transition on the falling 
edge of the clock (samples taken at instant `B' in Fig.~\ref{fig:alex_pd_timing}). 
Ideally, this sample results in the flip-flop `$F_2$' becoming metastable and resolving to 
`$0$' and `$1$' with equal probability to keep the loop locked.
It must be noted that in order to sample the data with the right
clock edge at the center of the data eye, the phase detector aligns the opposite edge 
with the data transitions.

\begin{figure}[h!] 
\centering 
\psfrag{A}{A} 
\psfrag{B}{B} 
\psfrag{C}{C}
\psfrag{vth}{$V_{th}$}
\includegraphics[width=6cm]{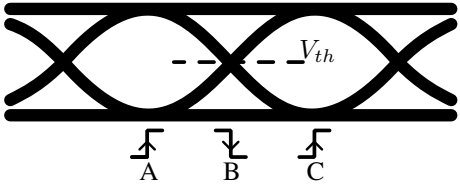} 
\caption{Sketch of a timing diagram of sampling instants of an Alexander phase detector
when used at the receivers of high speed data links.} 
\label{fig:alex_pd_isi} 
\end{figure}

In high speed systems, the phase detector operates over data that has high rise
and fall times and, more often than not, over data that has ISI
resulting in a finite horizontal eye opening.  Fig.~\ref{fig:alex_pd_isi} shows
a sketch of an eye diagram illustrating such a case.
In this case, for the loop to remain locked the sample taken at instance `B'
should resolve to `$0$' and `$1$' with equal likelihood. Note that the value that sample 
`B' resolves to depends on the jitter in the data and the input offset of the
sampling flip-flop `$F_2$'.
Since the clock phase corrections depend on the value of sample `B', the jitter in the 
recovered clock depends on the sequence of values that `B' resolves to.

%

\section{Analysis of the effect of sampler offset on data dependent jitter.}
\label{sec:jitteroffset}
The jitter in the data has two main components, which are random jitter (RJ) and 
data dependent jitter (DDJ). The DDJ comes from ISI, and is generally the dominant 
source of jitter~\cite{kundert_CDR,ddj_prediction}. The amount of DDJ depends on 
the bandwidth of the channel through which the data is being received.  

\begin{figure}[h!]
\centering
\psfrag{Data}{\small{Data}}
\psfrag{Clock}{\small{Clock}}
\psfrag{DN}{\small{DN}}
\psfrag{UP}{\small{UP}}
\psfrag{D}{\small{D}}
\psfrag{Q}{\small{Q}}
\psfrag{D1}{\small{$F_1$}}
\psfrag{D2}{\small{$F_2$}}
\psfrag{S}{\small{$\Sigma$}}
\psfrag{VCO}{\small{VCO}}
\psfrag{Charge}{\small{Charge}}
\psfrag{pump}{\small{pump}}
\psfrag{offset}{\small{V$_\text{off}$}}
\psfrag{R}{\small{R}}
\psfrag{C}{\small{C}}
\includegraphics[width=\columnwidth]{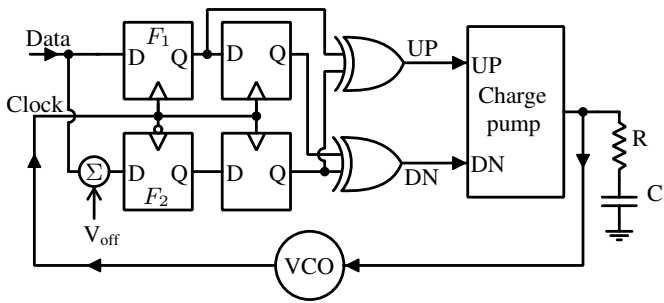}
\caption{Circuit schematic of a clock data recovery system.}
\label{fig:cdr}
\end{figure}

To analyze the effect of jitter in the data, we use VerilogA model of a simple 
clock recovery circuit as shown in Fig.~\ref{fig:cdr}. 
To simulate the effect of offset of flip-flop `$F_2$', different 
amounts of offset V$_\text{off}$ is added to the `D' input of `$F_2$' as shown in 
Fig.~\ref{fig:cdr}.
In order to simulate different amounts of DDJ, a 20 
section RLC network with different time constants is used. 

\subsection{Case 1: High bandwidth channel}
\label{subsec:hbw}
\begin{figure}[h!]
\centering
\psfrag{Data}{\small{Data (mV)}}
\psfrag{TwoUI}{\small{time (UI)}}
\psfrag{jitter}{\hspace{-2ex}\small{jitter$_\text{pk-pk}$ (\%UI)}}
\psfrag{offset}{\small{V$_\text{off}$~(mV)}}
\subfloat[Eye diagram for a high bandwidth channel. Inset plot shows histogram of 
zero crossing locations. Region demarcated by the red dashed box is expanded in 
part (b).\label{fig:d-eye-0bit}]
{\includegraphics[width=0.85\columnwidth]{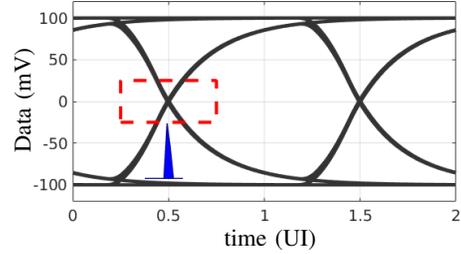}} \\
\subfloat[Zoomed view of the zero crossing region of the data is shown on the plot to 
the left. The plot to the right is the peak to peak jitter of the recovered clock 
vs offset of the sampling flip-flop `$F_2$'. Note that the plot to the right has the 
independent variable (offset V$_\text{off}$) on the y-axis because it is aligned to the 
voltage corresponding to the data on the plot to the left.\label{fig:jitt_0bit}]{
\includegraphics[width=0.8\columnwidth]{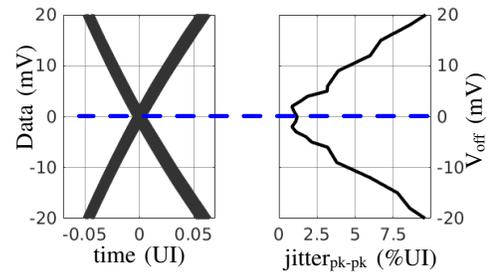}}
\caption{Eye diagram at the receiver and the corresponging jitter in the recovered clock 
vs offset voltage.}
\label{fig:benign}
\end{figure}

When the channel has reasonably good bandwidth, the DDJ is low and 
Fig.~\ref{fig:benign}\subref{fig:d-eye-0bit} shows the data eye diagram 
for such a case.
The inset plot in Fig.~\ref{fig:benign}\subref{fig:d-eye-0bit} shows the distribution of the zero crossing 
time instants of the data. It can be seen that the jitter histogram of the data is 
tightly distributed around a single peak. 
This data is used in the clock recovery circuit shown in Fig.~\ref{fig:cdr}, 
and simulated for different values of offset V$_\text{off}$. The peak-to-peak 
jitter of the recovered clock is plotted as a function of the offset in 
Fig.~\ref{fig:benign}\subref{fig:jitt_0bit}. It is perhaps not surprising that 
the jitter is minimum when the input offset of the flip-flop `$F_2$' is zero.

\subsection{Case 2: Moderate bandwidth channel}
\label{subsec:modbw}
\begin{figure}[h!]
\centering
\psfrag{Data}{\small{Data (mV)}}
\psfrag{TwoUI}{\small{time (UI)}}
\psfrag{jitter}{\small{jitter$_\text{pk-pk}$ (\%UI)}}
\psfrag{offset}{\small{V$_\text{off}$~(mV)}}
\subfloat[Eye diagram for a channel which results in ISI. Inset plot shows histogram of
zero crossing locations. Region demarcated by the red dashed box is expanded in
part (b).\label{fig:d-eye-1bit}]{
\includegraphics[width=0.9\columnwidth]{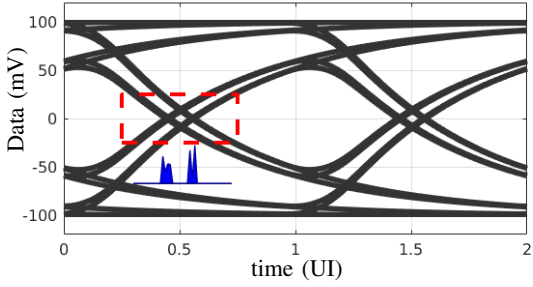}} \hspace{2ex}
\subfloat[Zoomed view of the zero crossing region of the data is shown on the plot to
the left. The plot to the right is the peak to peak jitter of the recovered clock
vs offset of the sampling flip-flop `$F_2$'. Note that the plot to the right has the
independent variable (offset V$_\text{off}$) on the y-axis because it is aligned to the
voltage corresponding to the data on the plot to the left.\label{fig:jitt_1bit}]{
\includegraphics[width=0.8\columnwidth]{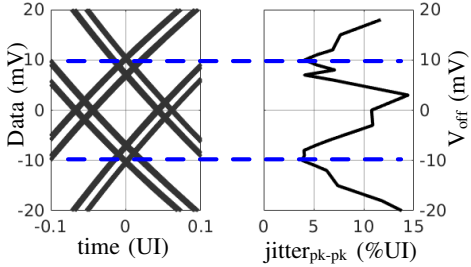}}
\caption{Eye diagram at the receiver for a channel with ISI and the corresponding 
jitter in the recovered clock vs offset voltage.}
\label{fig:harsh}
\end{figure}

\begin{figure}[h!]
\centering
\psfrag{B}{Q}
\psfrag{A}{P}
\psfrag{R}{R}
\psfrag{alpha}{$\cA$}
\psfrag{beta}{$\cB$}
\psfrag{gamma}{$\cC$}
\psfrag{b-1}{\small{$b[-1]$}}
\psfrag{b0}{\small{$b[0]$}}
\psfrag{b-2}{\small{$b[-2]$}}
\psfrag{bm}{\small{$b_m$}}
\psfrag{t0}{$\tau_0$}
\psfrag{t1}{$\tau_1$}
\psfrag{t2}{$\tau_2$}
\psfrag{t3}{$\tau_3$}
\psfrag{t4}{$\tau_4$}
\psfrag{s1}{\small{$X_1$}}
\psfrag{s2}{\small{$X_2$}}
\psfrag{s3}{\small{$X_3$}}
\psfrag{s4}{\small{$X_4$}}
\includegraphics[width=0.75\columnwidth]{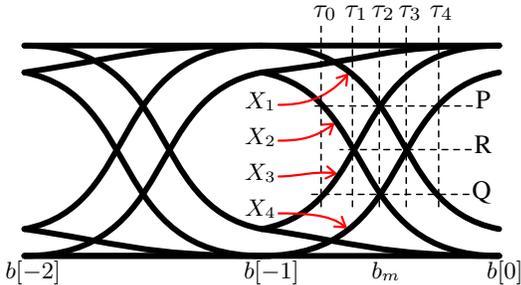}
\caption{Sketch of an eye diagram with strictly 1-bit ISI. $b[-2]$, 
$b[-1]$, and $b[0]$ are samples of the data taken on consecutive 
rising edges of the clock and $b_m$ is sample of data taken on 
the falling edge.
Nominal 
locations are shown.}
\label{fig:eye-1bit}
\end{figure}

When the channel bandwidth reduces the ISI in the data increases, resulting 
in higher DDJ, and such a case is shown in Fig.~\ref{fig:harsh}\subref{fig:d-eye-1bit}.
The inset plot in Fig.~\ref{fig:harsh}\subref{fig:d-eye-1bit} 
shows the distribution of the zero crossing
time instants of the data. In this case, the jitter histogram has two peaks, indicating that 
the ISI is dominated by 1 previous bit\footnote{Lower bandwidth cases can be approximated 
in a similar way with 2 or more bits.}~\cite{jitter-ber}.
When this data is used in the clock recovery circuit of Fig.~\ref{fig:cdr}
and simulated for different values of offset V$_\text{off}$, it is seen that 
the minimum jitter in the recovered clock is \emph{not} when the offset 
is zero. The peak-to-peak jitter of the recovered clock is plotted in 
Fig.~\ref{fig:harsh}\subref{fig:jitt_1bit} for different values of V$_\text{off}$. 
In this case, the jitter in the 
recovered clock is minimum when the offset is $\approx \pm 10$~mV.
The sampling threshold for minimum jitter is \emph{not} the voltage at which 
the spread in the transition times is the smallest. Note that the spread in the 
data transition times in least at 0~mV.
%
To understand why this is the case, we need to analyze the data sequences which 
result in corrective actions of increasing and reducing the clock frequency. 
Since the ISI is dominated by 1 previous bit, we can construct 
an approximate model of strictly 1-bit ISI and use it for analysis~\cite{nav-spring18}. 
Fig.~\ref{fig:eye-1bit} shows the sketch of an eye diagram where the ISI is limited 
to exactly 1 previous bit.
Here $b[-2]$, $b[-1]$ and $b[0]$ are three most recent bits. The Alexander phase detector 
uses samples $b[-1]$, $b_m$ (taken on the clock \FE) and $b[0]$ for 
detecting the phase difference between the data and the clock.
If the data comes from an equi-probable source, the data traces $X_1$, $X_2$, $X_3$ 
and $X_4$ in Fig.~\ref{fig:eye-1bit} occur with an equal probability of $1/8$ 
each~\cite{nav-spring18}. 
Let us now consider a few cases for the sampling threshold of `$F_2$'.
\begin{itemize}
\item {\bf Sampling threshold at `R' and clock \FE ~is between $\boldsymbol{\tau_1}$ and 
$\boldsymbol{\tau_3}$:}  In this case, the Alexander phase detector 
produces an `UP' signal for every trace $X_1$ and $X_4$ and a `DN' signal for every 
trace $X_2$ and $X_3$. Thus, the falling edge of the clock drifts around between $\tau_1$ and 
$\tau_3$.
\item {\bf Sampling threshold at `Q':} In this case, the phase detectors 
decision depends on the location of the clock. 
\begin{itemize}
\item {\bf Clock \FE ~between $\boldsymbol{\tau_0}$ and $\boldsymbol{\tau_2}$:} Here, traces $X_1$, $X_2$ and $X_4$ 
produce a `DN' signal and only $X_3$ produces an `UP' signal, pushing the clock \FE ~towards $\tau_2$.
\item {\bf Clock \FE ~between $\boldsymbol{\tau_2}$ and $\boldsymbol{\tau_4}$:} Here, traces $X_2$, $X_3$ and $X_4$
produce a `UP' signal and only $X_1$ produces an `DN' signal, pulling the clock \FE ~towards $\tau_2$.
\end{itemize}
Collectively these result in the clock's trailing edge being kept at $\tau_2$, thus minimizing the jitter.
\end{itemize}
One can similarly work out any other threshold value. The threshold levels `P' and `Q' 
are the levels where the jitter in the recovered clock is minimum. In the next section, 
we discuss the design of a circuit that automatically tracks the offset to keep it at 
the minimum jitter point.

\section{Automatic tracking of sampling threshold}
\label{sec:auto}
In the previous section we have seen that the jitter in the 
recovred clock is minimum when the sampling threshold is at `P' or at `Q'. By detecting 
the data sequence, it is possible to generate a circuit that adjusts the offset to bring it 
to a desired level. In this section, we will design a circuit that tunes the sampling 
threshold to keep it at `Q'.

\begin{figure}[h!]
\centering
\psfrag{Off\_UP}{UP$_{\text{V}_\text{th}}$}
\psfrag{Off\_DN}{DN$_{\text{V}_\text{th}}$}
\psfrag{DN}{DN}
\psfrag{UP}{UP}
\psfrag{ck}{ck}
\psfrag{ckb}{$\overline{\text{ck}}$}
\psfrag{Data}{Data}
\psfrag{D}{D}
\psfrag{D1}{D$_\text{1}$}
\psfrag{D2}{D$_\text{2}$}
\psfrag{Q}{Q}
\psfrag{S}{$\boldsymbol\Sigma$}
\includegraphics[width=\columnwidth]{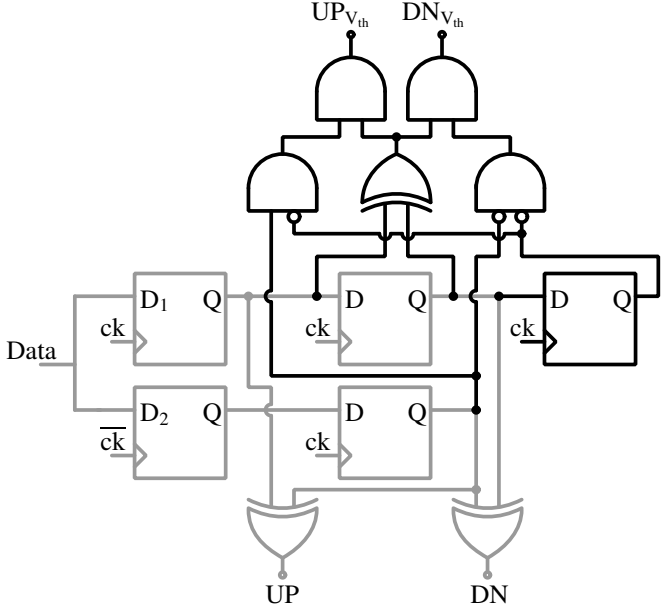}
\caption{Circuit for detecting whether the sampling threshold should be increased 
or decreased for reducing the jitter in the recovered clock. Part drawn in gray 
is the Alexander phase detector.}
\label{fig:new_pd}
\end{figure}

In order to adjust the sampling threshold and bring it to `Q', we need to identify the 
4 traces $X_1$ through $X_4$ and the sampling threshold. This can easily be done by looking at the three most recent bits 
received at the receiver, which are $b[-2]$, $b[-1]$ and $b[0]$. The Alexander phase detector 
also samples the data at the falling edge of the clock. let this bit be $b_m$. This sequence 
of 4 bits $b[-2] b[-1] b_m b[0]$ can be used to identify the location of the sampling 
threshold. This is tabulated in Table~\ref{tbl:truthtable}, wherein the threshold location 
and the action to be taken so as to bring the sampling threshold to `Q' are also tabulated.
\begin{table}[h!]
\begin{center}
\small{
\caption{Possible sequences and threshold locations for data with 1 bit ISI (refer Fig.~\ref{fig:eye-1bit})}
\label{tbl:truthtable}
\begin{tabular}{|c|c|c|c||c|c|}
\hline
$b[-2]$ & $b[-1]$ & $b_{m}$ & $b[0]$ & \multirow{2}{1.5cm}{\centering Threshold region} & \multirow{2}{*}{Action} \\
\RE & \RE & \FE & \RE & & \\ \hline
0 & 0 & 0 & 0 & $\cX$ & $NA$ \\ \hline
\RBT{0} & \RBT{0} & \RBT{0} & \RBT{1} & \RBT{$\boldsymbol{\text{V}_\text{th}>}$Q} & \OD \\ \hline
0 & 0 & 1 & 0 & $\cX$ & $NA$ \\ \hline
\RBT{0} & \RBT{0} & \RBT{1} & \RBT{1} & \RBT{$\boldsymbol{\text{V}_\text{th}<}$Q} & \OU \\ \hline
\RBT{0} & \RBT{1} & \RBT{0} & \RBT{0} & \RBT{$\boldsymbol{\text{V}_\text{th}>}$P OR $>$Q} & \OD \\ \hline
0 & 1 & 0 & 1 & $\cX$ & $NA$ \\ \hline
\RBT{0} & \RBT{1} & \RBT{1} & \RBT{0} & \RBT{$\boldsymbol{\text{V}_\text{th}<}$Q OR $<$P} & \OU \\ \hline
0 & 1 & 1 & 1 & $\cX$ & $NA$ \\ \hline
1 & 0 & 0 & 0 & $\cX$ & $NA$ \\ \hline
1 & 0 & 0 & 1 & $\text{V}_\text{th}>$P OR $>$Q & * \\ \hline
1 & 0 & 1 & 0 & $\cX$ & $NA$ \\ \hline
1 & 0 & 1 & 1 & $\text{V}_\text{th}<$Q OR $<$P& * \\ \hline
1 & 1 & 0 & 0 & $\text{V}_\text{th}>$P & * \\ \hline
1 & 1 & 0 & 1 & $\cX$ & $NA$ \\ \hline
1 & 1 & 1 & 0 & $\text{V}_\text{th}<$P & $NA$ \\ \hline
1 & 1 & 1 & 1 & $\cX$ & $NA$ \\ \hline
\end{tabular}}
\end{center}
$NA$: No action. $\cX$: No decision possible. \\ 
*Decision possible, but not used in this implementation.
\end{table}
From Table~\ref{tbl:truthtable}, we can find the logic condition for detecting when the 
sampling threshold should be increased and when it should be decreased. The expressions are 
\begin{align*}
\text{UP}_{\text{V}_\text{th}} = \overline{b[-2]}~\cdot~{b_m}~\cdot~(b[-1] \oplus b[0])  \\
\text{DN}_{\text{V}_\text{th}} = \overline{b[-2]}~\cdot~\overline{b_m}~\cdot~(b[-1] \oplus b[0]) 
\end{align*}
The logic circuit implementation of above expressions is shown in Fig.~\ref{fig:new_pd}. 
The parts drawn in gray in Fig.~\ref{fig:new_pd} form the Alexander phase detector.

\begin{figure}[h!]
\centering
\psfrag{D1}{D$_\text{1}$}
\psfrag{D2}{D$_\text{2}$}
\psfrag{DN}{\small{DN}}
\psfrag{UP}{\small{UP}}
\psfrag{UPof}{\small{UP$_{\text{V}_\text{th}}$}}
\psfrag{DNof}{\small{DN$_{\text{V}_\text{th}}$}}
\psfrag{Clock}{\small{Clock}}
\psfrag{Data}{\small{Data}}
\psfrag{S}{$\boldsymbol\Sigma$}
\psfrag{Phase}{\small{Phase}}
\psfrag{detector}{\small{detector}}
\psfrag{from }{\small{from}}
\psfrag{Fig}{\small{Fig.~\ref{fig:new_pd}}}
\psfrag{offset}{\small{V$_\text{th}^\text{fb}$}}
\psfrag{Charge}{\small{Charge}}
\psfrag{pump}{\small{pump}}
\psfrag{VCO}{VCO}
\psfrag{R}{\small{R}}
\psfrag{C}{\small{C}}
\psfrag{vc}{\small{V$_\text{c}$}}
\includegraphics[width=0.8\columnwidth]{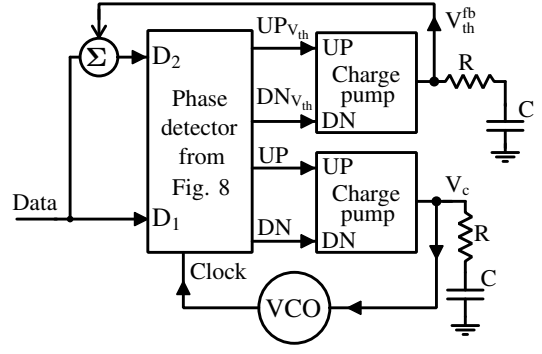}
\caption{Clock data recovery circuit that recovers sampling clock as well as optimum 
sampling threshold.}
\label{fig:cdr_LJ}
\end{figure}

\begin{figure}[h!]
\centering
\psfrag{vc}{\hspace{-5ex} \small{V$_\text{c}$~(mV)}}
\psfrag{vb}{\hspace{-5ex} \small{V$_\text{th}^\text{fb}$~(mV)}}
\psfrag{time}{\small{time ($\mu$s)}}
\includegraphics[width=.9\columnwidth]{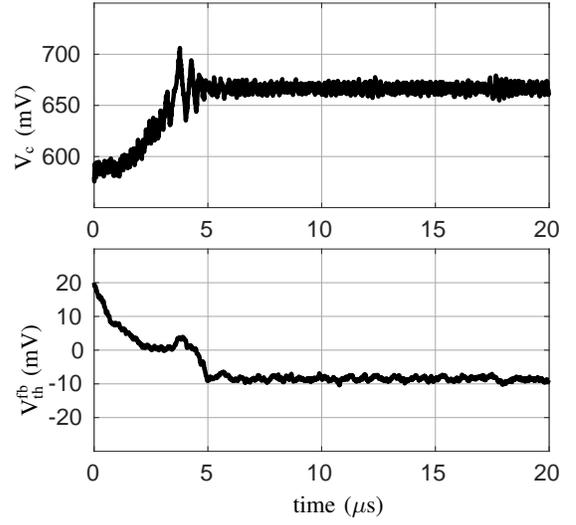}
\caption{Time evolution of control voltage (V$_\text{c}$) for VCO and that of 
sampling threshold (V$_\text{th}^\text{fb}$).}
\label{fig:vcvoff}
\end{figure}

This circuit, which senses the phase error as well as 
the threshold error, is used in a clock clock data recovery circuit as  
shown in Fig.~\ref{fig:cdr_LJ}. 
This circuit was simulated to verify its working for clock recovery and for sampling threshold 
recovery with data coming from a channel discussed in \emph{Case 2} in 
Section~\ref{subsec:modbw}. Fig.~\ref{fig:vcvoff} shows the time evolution of the VCO 
control voltage and sampling threshold V$_\text{th}^\text{fb}$. As expected 
V$_\text{th}^\text{fb}$ settles to $\approx-10$~mV, which corresponds to `Q'.

It can be shown that the same logic circuit recovers the optimum sampling threshold 
even for high bandwidth channels discussed in \emph{Case 1} in Section~\ref{subsec:hbw}.
It may be noted that this analysis and the results can also be used to design 
low jitter clock recovery circuits for PAM4 signals.

\section{Conclusions}
\label{sec:conclusion}
We have shown that input offset of the sampling flip-flops in the Alexander phase detector 
can influence the jitter in the recovered clock. We have shown that the effect of sampler offset on the 
recovered clock jitter also depends on the amount of ISI present in the data input. 
Using the fact that immediate previous bits are the highest 
contributors to ISI, we have constructed an approximate model assuming 1-bit ISI. Using this model 
we have explained the dependence of recovered clock jitter on input offset and also used this analysis to 
construct a circuit that recovers the optimum sampling threshold. The circuits are 
verified using simulations.

\bibliographystyle{IEEE}
\bibliography{IEEEabrv,refs}
\end{document}